
\input phyzzx
\input epsf
\font\Biggbf=cmbx10 scaled \magstep4
\def\half{{1\over{2}}}
\def\dslash{\not{\hbox{\kern-2pt $\partial$}}}
\def\bpsi{ {\overline \psi} }

\def\be{\beta}

\def\de{\delta}
\def\ep{\epsilon}

\def\et{\eta}

\def\la{\lambda}
\def\th{\theta}

\def\si{\sigma}

\def\Ga{\Gamma}
\def\La{\Lambda}
\def\De{\Delta}
\date{September, 1995}
\titlepage
\title{\Biggbf  Exact Boundary Scattering Matrices of the Supersymmetric
           Sine-Gordon Theory on a Half Line}
\author{Changrim Ahn$^{1}$
\foot{E-mail address: ahn@ewhahp3.ewha.ac.kr}
and Wai-Ming Koo$^{2}$
\foot{E-mail address: wkoo@Enterprise.snu.ac.kr}   }
\address{$^1$Department of Physics \break Ewha Womans University
			 \break  Seoul 120-750, S. Korea}
\address{$^{2}$Center for Theoretical Physics \break
Seoul National University \break Seoul 151-742, S. Korea}
\abstract{
Using the boundary Yang-Baxter equations and
exact results on the bulk $S$-matrices,
we compute exact boundary scattering amplitudes
of the supersymmetric sine-Gordon model with integrable boundary potentials.
   }
\endpage


\REF\ZamZam{A. Zamolodchikov and Al. Zamolodchikov,
Annal. Phys. {\bf 120} (1979)  253.}
\REF\GhoZam{S. Ghoshal and A. Zamolodchikov, Int. J. Mod. Phys. {\bf A9}
(1994) 3841.}
\REF\OliWit{E. Witten and D. Olive, Phys. Lett. {\bf B78} (1978)  97.}

{\bf 1. Introduction}

Quantum integrable models in two dimensions have been actively studied
since we can learn many aspects of non-perturbative physics which can
not be realized in more realistic models.
These models are providing arena where one can test new ideas
and get meaning of nontrivial solutions.
In addition these models do provide realistic models in a limited number
of applications.
The quantum sine-Gordon (SG) model is the most well-known exactly solvable
interacting quantum field theory.
Due to many theoretical developments, we now understand many nonperturbative
behaviors of this model.
Among them is the exact $S$-matrix $S_{\rm SG}$ of the solitons(+) and
antisolitons(-) given by\refmark\ZamZam:
$$\eqalign{&S^{++}_{++}(\th)=S^{--}_{--}(\th)=U(\th)
\sinh\left[\la(i\pi-\th)\right]  \cr
&S^{+-}_{-+}(\th)=S^{-+}_{+-}(\th)=
U(\th)\sinh(i\pi\la),\quad
S^{+-}_{+-}(\th)=S^{-+}_{-+}(\th)=
U(\th)\sinh(\la\th),\cr}   \eqn\Ssg$$
where $U(\th)$ is defined by
$$\eqalign{U(\th)&={1\over{i\pi}}\Ga(\la)\Ga\left(1+i{\la\th\over{\pi}}\right)
\Ga\left(1-\la-i{\la\th\over{\pi}}\right)
\prod^{\infty}_{l=1}{F_l(\th)F_l(i\pi-\th)\over{F_l(0)F_l(i\pi)}},\cr
F_l(\th)&={\Ga\left(2l\la+i{\la\th\over{\pi}}\right)
\Ga\left(1+2l\la+i{\la\th\over{\pi}}\right)\over{
\Ga\left((2l+1)\la+i{\la\th\over{\pi}}\right)
\Ga\left(1+(2l-1)\la+i{\la\th\over{\pi}}\right) }},\cr}\eqn\utheta$$
where $\la={8\pi\over{\be^2}}-1$ with $\be$ a usual coupling constant.

Recently there has been a lot of developments in the study of the
integrable models on the half-line or other restricted domain of the 1D space.
The main motivation is that these models can be applied to 3D spherically
symmetric physical systems where $s$-wave element becomes dominant.
One-channel Kondo problem, monopole-catalyzed proton decay are
frequently cited examples of these.
The quantum SG model on the half-line preserves the integrability
if the boundary potential is given by\refmark\GhoZam\
$${\cal L}_{\rm b}=\La\cos\left({\be(\phi-\phi_{0})\over{2}}\right).\eqn\Bsg$$
The integrability makes it possible to describe this theory
as a factorizable scattering theory of the solitons and their bound states.

Due to the existence of the boundary, we should introduce one more
scattering amplitude, the boundary scattering amplitudes $R_{a}^{b}(\th)$,
in addition to the bulk scattering.
For the bulk, the multi-particle scattering amplitudes are
factorized into a product of two-particle $S$-matrices if they
satisfy the Yang-Baxter equation (YBE) as a consistency condition.
For the boundary scattering, where particles scatter off with the boundary,
we need a new type of consistency condition, namely, the boundary
Yang-Baxter equation (BYBE) (also known as the reflection equation),
which can be expressed as:
$$\eqalign{
&\sum_{c_1,c_2,d_1,d_2} S^{c_{1}c_{2}}_{a_{1}a_{2}}(\th_1-\th_2)
R^{d_{1}}_{c_{1}}(\th_{1})
S^{d_{2}b_{1}}_{c_{2}d_{1}}(\th_1+\th_2)R^{b_{2}}_{d_{2}}(\th_2)\cr
&=\sum_{c_1,c_2,d_1,d_2}
R^{c_{2}}_{a_{2}}(\th_2)S^{c_{1}d_{2}}_{a_{1}c_{2}}(\th_1+\th_2)
R^{d_{1}}_{c_{1}}(\th_2)S^{b_{2}b_{1}}_{d_{2}d_{1}}(\th_1-\th_2).}\eqn\bybei$$

Besides the YBE, we need the unitarity and crossing-symmetry requirements
to fix the scattering amplitudes completely up to CDD ambiguity.
For the boundary scattering, this condition is expressed as the boundary
cross-unitarity condition,
$$R_{\bar a}^{b}\left({i\pi\over{2}}-\th\right)=S_{a'b'}^{ab}(2\th)
R_{\bar b'}^{a'}\left({i\pi\over{2}}+\th\right).\eqn\bcu$$
For later use, we summarize the known results of the boundary SG model
($A,\ {\overline A}$ and $B$ stand for soliton, antisoliton, and the boundary
respectively):
$$\eqalign{
&A(\th)B=P_{+}(\th)A(\th)B+Q_{+}(\th){\overline A}(\th)B,\quad
{\overline A}(\th)B=P_{-}(\th){\overline A}(\th)B+Q_{-}(\th)A(\th)B,\cr
&P_{\pm}(\th)=\cos(\xi\mp i\la\th)R(u),\quad
Q_{\pm}=-{k\over{2}}\sin(2i\la\th)R(u),}\eqn\Rsgi$$
where the prefactor is given by $R(\th)=R_{0}(\th)R_{1}(\th)$ with
$$\eqalign{
&R_{0}(\th)={\Ga\left(1+2i{\la\th\over{\pi}}\right)
\Ga\left(\la-2i{\la\th\over{\pi}}\right)\over{
\Ga\left(1+2i{\la\th\over{\pi}}\right)
\Ga\left(\la-2i{\la\th\over{\pi}}\right)}}
\prod^{\infty}_{k=1}{F_{2k}(2\th)\over{F_{2k}(-2\th)}}\cr
&R_{1}(\th)={1\over{\cos\xi}}\si(\eta,-i\th)\si(i\vartheta,-i\th),}\eqn\Rsgii$$
where $F_{k}$ is given in Eq.\utheta\ and
$$\eqalign{
&\si(x,u)={\Pi\left(x,{\pi\over{2}}-u\right)\Pi\left(-x,{\pi\over{2}}-u\right)
\Pi\left(x,-{\pi\over{2}}+u\right)\Pi\left(-x,-{\pi\over{2}}+u\right)\over{
\Pi^{2}\left(x,{\pi\over{2}}\right)\Pi^{2}\left(-x,{\pi\over{2}}\right)}},\cr
&\Pi(x,u)=\prod_{l=0}^{\infty}
{\Ga\left[\half+\left(2l+\half\right)\la+{x\over{\pi}}-{\la u\over{\pi}}\right]
\Ga\left[\half+\left(2l+{3\over{2}}\right)\la+{x\over{\pi}}\right]\over{
\Ga\left[\half+\left(2l+{3\over{2}}\right)\la+{x\over{\pi}}
-{\la u\over{\pi}}\right]\Ga\left[\half+\left(2l+\half\right)\la
+{x\over{\pi}}\right]}}.         }\eqn\Rsgiii$$
The parameters $\et,\vartheta$ are related to $k,\xi$ by
$$\cos\et\cos\vartheta=-{1\over{k}}\cos\xi,\quad
\cos^{2}\et+\cos^{2}\vartheta=1+{1\over{k^{2}}}.\eqn\para$$
The relationship between these parameters and those in Eq.\Bsg\
$M,\phi_{0}$ is not completely understood.\refmark\GhoZam\

\REF\DiVFer{P. Di Vecchia and S. Ferrara, Nucl. Phys. {\bf B130} (1977)  93.}
\REF\Chaichian{M. Chaichian and P. Kulish, Phys. Lett. {\bf B183} (1987) 169;
O. Babelon and F. Langouche, Nucl. Phys. {\bf B290} [FS20] (1987) 603;
M.A. Olshanesky, Comm. Math. Phys. {\bf 88} (1983) 63.}
\REF\Inami{T. Inami, S. Odake, and Y.-Z. Zhang, preprint {\tt hep-th/9506157}
Supersymmetry Extension of the Sine-Gordon Theory with Integrable Boundary
Interactions.}
\REF\BerLeC{D. Bernard and A. LeClair, Nucl. Phys. {\bf B340} (1990) 721.}
\REF\ResSmi{N. Yu Reshetikhin and F. Smirnov, Comm. Math. Phys. {\bf 131}
(1990) 157.}
\REF\ABL{C. Ahn, D. Bernard and A. LeClair, Nucl. Phys. {\bf B346} (1990) 409.}
\REF\Ahn{C. Ahn, Nucl. Phys. {\bf B354} (1991) 57.}

{\bf 2. Supersymmetric Sine-Gordon Theory without Boundary}

Our objective in this paper is to solve the $N=1$ SUSY
sine-Gordon (SSG) theory on the half-line with an appropriate boundary
potential which preserves the integrability.
The action of the SSG model is given by\refmark\DiVFer\
$$ S= \int dx dt \left[ \half(\partial_{\mu}\phi)^2
-i\bpsi\dslash\psi - {m^2\over{\be^2}}\cos^2\phi
- 2m(\cos{\be\phi\over{2}})\bpsi\psi \right],\eqn\ssgaction$$
where $\phi$ is a real scalar field and  $\psi$ is a Majorana fermion.
$\beta$ is a coupling constant of the SG theory and $m$ is the mass
parameter denoting the deviation from the massless theory.
The SSG theory is integrable because it is equivalent to Toda theory on
the twisted super affine Lie algebra $C^{(2)}(2)$.\refmark\Chaichian\
The equation of motion of the SSG theory can be written as a
super zero-curvature condition.
An infinite number of conserved charges at the classical level were derived
and seem to be preserved at the lowest order of quantum level.
Due to the integrability, there exist solitons and anti-solitons, as
well as their bound states. All these particles form a SUSY multiplet.
In this model the SUSY algebra is extended by the central charge which
is the topological charge of the soliton and antisoliton.\refmark\OliWit

Exact results on the SSG theory has been derived due to the development
of the perturbed CFTs.
It has been well-known that the $S$-matrix of the minimal CFTs
${\cal M}_{p/p+1}$ with the central charge $c=1-{6\over{p(p+1)}}$
perturbed by the least relevant operator can be obtained from the SG theory
by restricting the $S$-matrix into the RSOS type using the hidden
quantum group symmetry.\refmark{\BerLeC,\ResSmi}\
The particle spectrum of this so-called restricted SG (RSG) theory is
composed of the kinks $K_{ab}$ which connect two adjacent spins $a,b$
with $|a-b|=\half$, instead of the (anti)solitons.

The $S$-matrix of the RSG theory, $S^{(p)}_{\rm RSG}$, is given by
$$S^{ab}_{dc}(\th)=U(\th)\left(X^{ab}_{cd}\right)^{-{\th\over{2\pi i}}}
\left[\sqrt{X^{ab}_{cd}}\sinh\left({\th\over{p}}\right)\delta_{db}+
\sinh\left({i\pi-\th\over{p}}\right)\delta_{ac}\right],\eqn\Srsg$$
for the process $\vert K_{da_{1}}(\th_1)\rangle+\vert K_{a_{2}b}(\th_2)\rangle
\to\vert K_{dc_{1}}(\th_2)\rangle +\vert K_{c_{2}b}(\th_1)\rangle$
where
$X^{ab}_{cd}=\left({[2a+1][2c+1]\over{[2d+1][2b+1]}}\right)$ with
$q$-number $[n]=(q^{n}-q^{-n})/(q-q^{-1})$ and $q=-e^{-i\pi/p}$.
The allowed values of spins are $0,\half,1,...,{p\over{2}}-1$.

The exact $S$-matrix of the SSG theory has been obtained as a byproduct
from the result of the perturbed superCFT; the perturbed super CFTs have
the $S$-matrix in the factorized form of
$$S_{\rm SCFT}(\th)=S^{(4)}_{\rm RSG} (\th)\otimes
S^{(p)}_{\rm RSG}(\th),\eqn\scft$$
and by `unrestricting' $S^{(p)}_{\rm RSG}$, one obtains
the SSG (anti)soliton $S$-matrix\refmark\ABL:
$$S_{\rm SSG}(\th)=S^{(4)}_{\rm RSG} (\th)\otimes S_{\rm SG}(\th).\eqn\Sssg$$
The first $S$-matrix factor which commutes with the SUSY charges,
applies to the superspace indices of the
SSG particles and the second one is nothing but the SG (anti)soliton
$S$-matrix, Eq.\Ssg, however, with different parameterization,
$$\la={2\pi\over{\be^2}}-\half.\eqn\coupling$$

By denoting the SG solitons with topological charge
$\pm 1$ by $\vert A^{\pm}\rangle$,
the particle states of the SSG theory can be denoted by
$\left\vert K_{ab}^{\pm}\right\rangle=\vert K_{ab}\rangle\otimes\vert A^{\pm}
\rangle$, where the first quantum number carries the SUSY
charges and the second the topological charges.
Explicit SUSY transformations are as follows\refmark\Ahn:
$$\eqalign{
Q\left\vert K^{\pm}_{0\half}\right\rangle&=-i e^{\th/2}
\left\vert K^{\pm}_{1\half}\right\rangle,\qquad
{\overline Q}\left\vert K^{\pm}_{0\half}\right\rangle=\mp i
e^{-\th/2}\left\vert K^{\pm}_{1\half}\right\rangle,    \cr
Q\left\vert K^{\pm}_{1\half}\right\rangle&=\ \ i e^{\th/2}
\left\vert K^{\pm}_{0\half}\right\rangle,\qquad
{\overline Q}\left\vert K^{\pm}_{1\half}\right\rangle=\pm i
e^{-\th/2}\left\vert K^{\pm}_{0\half}\right\rangle,\cr} \eqn\susyi$$
and on the charge conjugated states by
$$\eqalign{
Q\left\vert K^{\pm}_{\half 0}\right\rangle
&=\ \ e^{\th/2}\left\vert K^{\pm}_{\half 0}\right\rangle,
\qquad {\overline Q}\left\vert K^{\pm}_{\half 0}\right\rangle=\pm e^{-\th/2}
\left\vert K^{\pm}_{\half 0}\right\rangle,\cr
Q\left\vert K^{\pm}_{\half 1}\right\rangle&= - e^{\th/2}
\left\vert K^{\pm}_{\half 1}\right\rangle,
\qquad {\overline Q}\left\vert K^{\pm}_{\half 1}\right\rangle
=\mp e^{-\th/2}\left\vert K^{\pm}_{\half 1}\right\rangle.\cr}\eqn\susyii$$
 From these relations, one can check that the SUSY charges satisfy
$$Q^2=P=e^{\th},\qquad {\overline Q}^2={\overline P}=e^{-\th},\qquad
{\rm and}\qquad Q{\overline Q}+{\overline Q}Q=2T.\eqn\Salgebra$$
The central charge $T$ is $\pm 1$ corresponding to the
topological charges of the SSG solitons.

{\bf 3. Boundary Scattering Matrices of the Supersymmetry Sine-Gordon Theory}

Integrability is often preserved by the introduction of the SUSY.
Therefore, one can naturally guess that the SUSY extension of the SG theory
on the half-line can be integrable.
In recent work, it has been claimed that the half-line SSG theory is
integrable if one introduces a well-defined boundary potential.
In ref.[\Inami], it has been shown that the SSG model can preserve
integrability with the following boundary potential:
$${\cal L}_{\rm b}=\La\cos\left({\be(\phi-\phi_{0})\over{2}}\right)
+M{\bar\psi}\psi+\ep\psi+{\bar\ep}{\bar\psi}.\eqn\Bssg$$
Due to this integrability, we can describe the boundary SSG model as
a scattering theory where the amplitudes can be obtained from the
BYBE.

Since the bulk SSG $S$-matrix has a factorized form,
we will restrict ourselves to find the boundary scattering $R$-matrix
in the factorized form
$$R_{\rm SSG}(\th)=R_{\rm SUSY}(\th)\otimes R_{\rm SG}(\th)\eqn\sol$$
as well.
So each factor satisfies the boundary YBE, Eq.\bybei separately.
The second factor is the usual SG part Eq.\Rsgi, with $\la$ given by
Eq.\coupling. The first factor is what we are going determine based on the
boundary YBE.

\REF\ShaWit{R. Shankar and E. Witten, Phys. Rev. {\bf D17} (1978) 2134.}

This boundary scattering matrix satisfies the boundary YBE in the RSOS
representation given by
$$\eqalign{
&\sum_{a_{1},b_{1}}R^a_{bb_{1}}(\th)S^{ac}_{b_{1}a_{1}}(\th_{2}+\th_{1})
R^{a_{1}}_{b_{1}b_{2}}(\th_{2})S^{a_{1}c}_{b_{2}a_{2}}(\th_{2}-\th_{1})\cr
&=\sum_{a_{1},b_{1}}S^{ac}_{ba_{1}}(\th_{2}-\th_{1})
R^{a_{1}}_{bb_{1}}(\th_{2})S^{a_{1}c}_{b_{1}a_{2}}(\th_{2}+\th_{1})
R^{a_{2}}_{b_{1}b_{2}}(\th_{1}),}\eqn\bybeii$$
where $R^{a}_{bc}(\th)$ denotes the boundary $S$-matrix and the bulk
scattering matrix $S^{ab}_{cd}(\theta)$ is given by Eq.\Srsg.

In general $R^{a}_{bc}(\th)$ contains both diagonal and off-diagonal
scattering components, which can be written as
$$R^{a}_{bc}(\th)=R(\th)\left(X^{bc}_{aa}\right)^{-{\th\over{2\pi i}}}
\left[\de_{b\neq c}X^{a}_{bc}(\th)+\de_{bc}\left(\de_{b-1/2,a}U_{a}(\th)
+\de_{b+1/2,a}D_{a}(\th)\right)\right],\eqn\bsm$$
where $R(\th)$ have to be determined from the boundary crossing and
unitarity constraints, while $X^{a}_{bc}$ and $U_a,D_a$ have to be
determined from the BYBE.  An overall q-number factor is
multiplied to the above to cancel that from the bulk $S$-matrix
in order to simplifies the BYBE.
For $p=4$, there are three RSOS spins labeled by $0,\half$, and $1$.
The unknown scattering amplitudes are $U_{0},D_1,U_{\half},D_{\half}$
for diagonal scattering weights and $X^{\half}_{01},X^{\half}_{10}$
for off-diagonal scattering weights.

Substituting Eq.\bsm\ into the BYBE, one finds that the
unknowns satisfy the following equations:
$$\eqalign{
&X^{\half}_{01}(\th')X^{\half}_{10}(\th)
=X^{\half}_{01}(\th)X^{\half}_{10}(\th')\cr
&U_{\half}(\th)(1+\sqrt{2}f_{-})+D_{\half}(\th')(1+\sqrt{2}f_{+})
(1+\sqrt{2}f_{-}) =U_{\half}(\th')+D_{\half}(\th)(1+\sqrt{2}f_{+})\cr
&D_{\half}(\th)(1+\sqrt{2}f_{-})+D_{\half}(\th')(1+\sqrt{2}f_{+})
=(1+\sqrt{2}f_{-}) D_{\half}(\th')+U_{\half}(\th)(1+\sqrt{2}f_{+})\cr
&U_0(\th')D_1(\th)f_{+}\left(1+{f_{-}\over{\sqrt{2}}}\right)
+D_1(\th')D_1(\th)f_{-}\left(1+{f_{+}\over{\sqrt{2}}}\right)\cr
&\qquad=U_0(\th)D_1(\th')f_{+}\left(1+{f_{-}\over{\sqrt{2}}}\right)
+U_0(\th')U_0(\th) f_{-}\left(1+{f_{+}\over{\sqrt{2}}}\right),}\eqn\sss$$
where
$$f_{-}={\sinh\left({\th'-\th\over{4}}\right)\over{
\sinh\left({i\pi-\th'+\th\over{4}}\right)}},\qquad
f_{+}={\sinh\left({\th'+\th\over{4}}\right)\over{
\sinh\left({i\pi-\th'-\th\over{4}}\right)}}.$$

It is important to point out that $U_{\half}$ ($U_0$) is coupled to
$D_{\half}$ ($D_1$) through the BYBE,
and the two equations relating $U_{\half},D_{\half}$
are there only if $X^{\half}_{01},X^{\half}_{10}$ are nonvanishing.
If the later are taken to be zero in the first place, i.e. off-diagonal
scattering is forbidden, then the BYBE does not provide any information
on $U_{\half},D_{\half}$, a case we will elaborate in the sequel.

 From Eq.\sss, we have
$$X^{\half}_{01}(\th)\propto X^{\half}_{10}(\th),\eqn\pro$$
the constant of proportionality can actually be shown to be a gauge factor,
hence difference between these two off-diagonal scattering amplitudes
is not significant at this point.
While the rest of the equations can be turned into ordinary
differential equations in the limit $\th'\to\th$,
giving respectively
$$\eqalign{
&\left[\dot{U}_{\half}(\th)+\dot{D}_{\half}(\th)\right]\tanh{\th\over{2}}
+2\left[U_{\half}(\th)+D_{\half}(\th)\right]=0\cr
&\left[\dot{U}_{\half}(\th)-\dot{D}_{\half}(\th)\right]\coth{\th\over{2}}
-2\left[U_{\half}(\th)-D_{\half}(\th)\right]=0,}\eqn\two$$
and
$$\dot{\cal R}(\th)\tanh{\th\over{2}}={\cal R}(\th)^2-1,\eqn\one$$
where
${\cal R}(\th)\equiv D_1(\th)/U_0(\th)$
and $\dot{\ \ }$ denotes differentiation with respect to $\th$.
These equations can be easily integrated to give
$$\eqalign{
&U_{\half}(\th)={B\over{\sinh{\th\over{2}}}}+C\cosh{\th\over{2}},\qquad
D_{\half}(\th)={B\over{\sinh{\th\over{2}}}}-C\cosh{\th\over{2}},\cr
&{D_1(\th)\over{U_0(\th)}}=
{1-A\sinh{\th\over{2}}\over{1+A\sinh{\th\over{2}}}},}\eqn\ratio$$
where $A,B,C$ are free parameters.

Next, we want to determine the overall $R(\theta)$ with
the use of the boundary unitarity and crossing symmetry conditions
which can be summarized as
$$\sum_{c}R^{a}_{bc}(\th)R^{a}_{cd}(-\th)=\de_{bd},\qquad
\sum_{d}S^{ac}_{bd}(2\th)R^{d}_{bc}\left({i\pi\over{2}}+\theta\right)
=R^{a}_{bc}\left({i\pi\over{2}}-\th\right).\eqn\eq$$

First we consider the common $R(\th)$ factor of the weights
$X^{\half}_{10},X^{\half}_{01},U_2,D_2$. The above two conditions give
respectively
$$R(\th)R(-\th)\left(X^{\half}_{01}X^{\half}_{10}+C^2\cosh^2{\th\over{2}}
-{B^2\over{\sinh^2{\th\over{2}}}}\right)=1,\eqn\ri$$
and
$$U(2\th)R\left({i\pi\over{2}}+\th\right)\sinh\left({i\pi\over{4}}
-{\th\over{2}}\right)
=R\left({i\pi\over{2}}-\th\right) . \eqn\rii$$
To arrive at the above used has been made of the following relations
$$\eqalign{
&D_{\half}(\th)=D_{\half}(i\pi-\th)\left[1+{\sqrt{2}\sinh
\left({i\pi-2\th\over{4}}\right) \over{\sinh{\th\over{2}} }}\right]\cr
&U_{\half}(\th)=U_{\half}(i\pi-\th)\left[1+{\sqrt{2}\sinh
\left({i\pi-2\th\over{4}}\right) \over{\sinh{\th\over{2}} }}\right],}\eqn\eq$$
which can be obtained from Eq.\sss,
taking the limit $\th'\to i\pi-\th$.
In the unitarity condition, the non-zero factors $X^{\half}_{01}$ and
$X^{\half}_{10}$ can be absorbed into $R(\th)$ and we set it as $-1$
for convenience.

\REF\MLSS{G. Mussardo, A. LeClair, H. Saleur, and S. Skorik,
preprint {\tt hep-th/9503227} Boundary energy and boundary states in integrable
quantum theories. }
To solve for $R(\th)$, we write
$$R(\th)=\sinh{\th\over{2}}R_0(\th)R_1(\th)$$
where now
$$R_{0}(\th)R_0(-\th)=1,\qquad
U(2\th)R_0\left({i\pi\over{2}}+\th\right)\sinh\left({i\pi\over{4}}
+{\th\over{2}}\right)
=R_0\left({i\pi\over{2}}-\th\right),\eqn\eqnon$$
whose minimal solution does not depend on the free parameters $B,C$.

While $R_1$ contains all the information of the boundary conditions and
satisfies
$$R_1(\th)R_1(-\th)
\left[B^2+(1-C^2)\sinh^2{\th\over{2}}-C^2\sinh^4{\th\over{2}}\right]=1,
\quad R_1(\th)=R_1(i\pi-\th).\eqn\r$$
The minimal solution is given by Eq.\Rsgii\ with the following replacements:
$$\cos^2\xi\to B^2,\quad k^2\to -C^2,\quad \la\to\half.$$

The $R(\th)$ factor of the weights $U_0,D_1$ need not be the same as
that determined above since as mentioned before these weights are not
coupled to those treated before. The unitarity condition gives
$$R(\th)R(-\th)U_0(\th) U_0(-\th)=1,\qquad R(\th)R(-\th)D_1(\th)D_1(-\th)=1.$$

While the crossing symmetry gives
$$\eqalign{
U(i\pi-2\th)R(i\pi-\th)&\left[
U_0(i\pi-\th)\sinh\left({2\th+i\pi\over{4}}\right)
+D_1(i\pi-\th)\sinh\left({i\pi-2\th\over{4}}\right)\right]\cr
&=\sqrt{2}R(\th)U_0(\th)\cr
U(i\pi-2\th)R(i\pi-\th)&\left[
D_1(i\pi-\th)\sinh\left({i\pi-2\th\over{4}}\right)
+U_0(i\pi-\th)\sinh\left({i\pi-2\th\over{4}}\right)\right]\cr
&=\sqrt{2}R(\th)D_1(\th).}\eqn\eq$$
We can solve these equations separately by requiring
$$R(\th)R(-\th)=1,\quad
U(2\th)R\left({i\pi\over{2}}+\th\right)\sinh\left({i\pi\over{4}}
-{\th\over{2}}\right)=R\left({i\pi\over{2}}-\th\right).\eqn\rfac$$
So that $R(\th)$ is given by $R_0(\theta)$ in Eq.\Rsgii\
with $\lambda={1\over{4}}$. While $U_0,D_1$ satisfy
$$U_0(\th) U_0(-\th)=1,\quad
D_1(\th) D_1(-\th)=1,\eqn\ui$$
and
$$\eqalign{
&U_0(i\pi-\th)\sinh\left({2\th+i\pi\over{4}}\right)+D_1(i\pi-\th)
\sinh\left({i\pi-2\th\over{4}}\right)=\sqrt{2} U_0(\th)\sinh{\th\over{2}},\cr
&D_1(i\pi-\th)\sinh\left({i\pi-2\th\over{4}}\right)+U_0(i\pi-\th)
\sinh\left({i\pi-2\th\over{4}}\right)=\sqrt{2} D_1(\th)\sinh{\th\over{2}}.}$$
These two sets of equations are compatible with Eqs.\ratio,\sss\
hence they contain only two independent relations.
Substituting the ratio of $D_1,U_0$ into the above we get a relation
between $U_0(\th)$ ($D_1(\th)$) and $U_0(i\pi-\th)$ ($D_1(i\pi-\th)$);
$$\eqalign{
&{U_0(\th)\over{U_0(i\pi-\th)}}={i\cosh{\th\over{2}}
\left(1+A\sinh{\th\over{2}}\right)\over{\sinh{\th\over{2}}
\left(1+iA\cosh{\th\over{2}}\right)}},\cr
&{D_1(\th)\over{D_1(i\pi-\th)}}={i\cosh{\th\over{2}}
\left(1-A\sinh{\th\over{2}}\right)\over{\sinh{\th\over{2}}
\left(1-iA\cosh{\th\over{2}}\right)}}.}$$
These relations together with Eq.\ui\ can determine
$U_0(\th)$  and $D_1(\th)$ up to the CDD factor;
$$\eqalign{
&U_0(\th)=\left({\sinh{i\De\over{2}}\over{\sinh{\th\over{2}}}}+1\right)
R(\th)R(i\pi-\th),\
D_1(\th)=\left({\sinh{i\De\over{2}}\over{\sinh{\th\over{2}}}}-1\right)
R(\th)R(i\pi-\th),\cr
&R(\th)={\Ga({-i\th\over{2\pi}})\over{\Ga(\half-{i\th\over{2\pi}})}}
\prod_{l=1}^{\infty}{\Ga({\De\over{2\pi}}-{i\th\over{2\pi}}+l)
\Ga(-{i\th\over{2\pi}}-{\De\over{2\pi}}+l-1)
\Ga^2(-{i\th\over{2\pi}}+l-\half)\over{
\Ga({\De\over{2\pi}}-{i\th\over{2\pi}}+l+\half)
\Ga(-{i\th\over{2\pi}}-{\De\over{2\pi}}+l-\half)
\Ga^2(-{i\th\over{2\pi}}+l-1)}},}$$
where $A^{-1}=i\sin{\De\over{2}}$.

Finally, we consider the case when $X^{\half}_{10},X^{\half}_{01}$ are
zero to begin with.
Then unitarity and crossing symmetry are the only conditions
that can be used to determined $U_{\half},D_{\half}$. These conditions give
$$R(\th)R(-\th)U_{\half}(\th) U_{\half}(-\th)=1,\qquad
R(\th)R(-\th)D_{\half}(\th)D_{\half}(-\th)=1.$$
and
$$\eqalign{
U(i\pi-2\th)R(i\pi-\th)U_{\half}(i\pi-\th)\sinh{i\pi-\th\over{2}}
&=R(\th)U_{\half}(\th)\cr
U(i\pi-2\th)R(i\pi-\th)D_{\half}(i\pi-\th)\sinh{i\pi-\th\over{2}}
&=R(\th)D_{\half}(\th).}$$

Again, we require separately
$R(\th)$ to satisfy the same relations given in Eq.\rfac\
and $U_{\half},D_{\half}$
to satisfy
$$U_{\half}(\th) U_{\half}(-\th)=1,\quad
D_{\half}(\th) D_{\half}(-\th)=1,\eqn\ud$$
and
$$\eqalign{
U_{\half}(i\pi-\th)\sinh{i\pi-\th\over{2}}
&=U_{\half}(\th)\sinh{\th\over{2}}\cr
D_{\half}(i\pi-\th)\sinh{i\pi-\th\over{2}}
&=D_{\half}(\th)\sinh{\th\over{2}}.}$$
 From these relations, we find
$U_{\half}(\th)=D_{\half}(\th)=i/[\sinh(\th/2)\si(\pi/2,-i\th)]$
where $\si$ is defined in \Rsgiii.

To summarize, we found two mutually exclusive sets of solutions for
$U_0$, $D_1$, $U_{\half}$, $D_{\half}$, $X^{\half}_{10}$, $X^{\half}_{01}$.
The first set has non-vanishing off-diagonal element ($X^{\half}_{10}\neq 0$)
and introduces the undetermined parameters $A,B,C$,
while the second set only allows diagonal scattering ($X^{\half}_{10}=0$)
with one free parameter $A$.
As for the number of free parameters, we should keep in mind the two more
parameters appearing in the SG soliton sector.
Therefore there are on total five or three parameters for the boundary
scattering theory. It should be remarked that our results are based on
the assumption that the boundary scattering matrix is in the factorized
form given in Eq.\sol, we only have proof that for the scattering amplitudes
that involve $U_0,D_1$ are indeed factorized. It is not known whether
there are other non-factorized scattering matrices that involve $U_{\half}$,
$D_{\half}$, $X^{\half}_{10}$ and $X^{\half}_{01}$.

%
%

{\bf 4. Discussion}

Our results on the scattering matrix suggest that there are at
least two integrable boundary lagrangian for the SSG model.
 From the number of parameters in the theory,  we claim that
the non-diagonal scattering theory
corresponds to the SSG model with the boundary potential given in Eq.\Bssg\
where five parameters are introduced. On making this claim, we have
assumed that there is no other integrable boundary lagrangian with
the five parameters. On the diagonal scattering theory with three parameters,
we do not know which is the boundary potential it corresponds to.

It is not clear in general how the five parameters in the scattering
theory and the lagrangian are related at this moment except for
a few special cases.
For $\La=\infty$ with $M,\ep,{\bar\ep}$ arbitrary, the topological
charge is preserved, therefore topological charge violating amplitudes
should be zero. Since our $R$-matrix has a factorized form in such a way
that SUSY sector is separated from the topological sector,
$Q_{\pm}=0$ ($k=0$) irrespective of the SUSY sector.
For $\La=0$ with $M,\ep,{\bar\ep}$ arbitrary, the charge conjugation
symmetry is preserved, and soliton and antisoliton irrespective of
their SUSY charges behave in the same way. This means $P_{+}=P_{-}$ and
$Q_{+}=Q_{-}$, or $\xi=0$.

For other special cases, it is more convenient to re-express the SUSY sector
in terms of the SUSY eigenstate,
$\vert B^{\pm}\rangle$ and $\vert F^{\pm}\rangle$
defined by
$$\eqalign{
&\vert B^{\pm}\rangle={1\over{\sqrt{2}}}\left(\vert K^{\pm}_{0\half}\rangle
+\vert K^{\pm}_{1\half}\rangle\right),\quad
\vert F^{\pm}\rangle={1\over{\sqrt{2}}}\left(\vert K^{\pm}_{0\half}\rangle
-\vert K^{\pm}_{1\half}\rangle\right),\cr
&\vert {\overline B}^{\pm}\rangle={1\over{\sqrt{2}}}
\left(\vert K^{\pm}_{\half 0} \rangle
+\vert K^{\pm}_{\half 1}\rangle\right),\quad
\vert {\overline F}^{\pm}\rangle={1\over{\sqrt{2}}}
\left(\vert K^{\pm}_{\half 0}\rangle -\vert K^{\pm}_{\half 1}\rangle\right).}$$
Now solitons and antisolitons carry well-defined fermion number $F$, say,
$F=0$ for $\vert B^{\pm}\rangle$ and $F=1$ for $\vert F^{\pm}\rangle$.
If we rewrite the boundary $R$-matrix in this basis, one finds
$$\eqalign{
&B^{\pm}(\th) B=
\half(D_{\half}+U_{\half}+X_{+}) B^{\pm}(-\th)B
+\half(D_{\half}-U_{\half}+X_{-}) F^{\pm}(-\th)B,\cr
&F^{\pm}(\th) B=
\half(D_{\half}-U_{\half}-X_{-}) B^{\pm}(-\th)B
+\half(D_{\half}+U_{\half}-X_{+}) F^{\pm}(-\th)B,\cr
&{\overline B}^{\pm}(\th) B=\half(U_{0}+D_{1}){\overline B}^{\pm}(-\th) B
+\half(U_{0}-D_{1}){\overline F}^{\pm}(-\th) B,\cr
&{\overline F}^{\pm}(\th) B=\half(U_{0}-D_{1}){\overline B}^{\pm}(-\th) B
+\half(U_{0}+D_{1}){\overline F}^{\pm}(-\th) B,}\eqn\eq$$
where $X_{\pm}=X^{\half}_{10}\pm X^{\half}_{01}$.

Now consider the case where $\ep={\bar\ep}=0$.
Since the lagrangian preserves fermion number,
the fermion number violating amplitudes in the $R$-matrix should vanish.
By choosing a gauge, we can fix $X^{\half}_{01}=X^{\half}_{10}$.
This leaves $D_{\half}=U_{\half}$ and $U_0=D_1$, which means
$A=C=0$ if $\ep={\bar\ep}=0$.
For more complete relations, it is desirable to consider the boundary
scattering of the SSG bound states since the lowest massive bound states
are $\phi$ and $\psi$ fields appearing in the lagrangian.
One can use the bootstrap procedure for this computation, although we
will not pursue it here.

It is interesting to note that the boundary SSG model introduces
extra boundary poles in addition to those from the SG model,
which in the physical strip may be interpreted as resonance states
in the supersymmetric theory with a boundary.

\ack

The work of CA is supported in part by KOSEF 941-0200-003-2 and
BSRI 94-2427 and that of WMK by a grant from KOSEF through SNU/CTP.
\endpage

\refout

\endpage

\end